\documentstyle[aps,epsf,prl,floats]{revtex}

\def\sb#1{$_{#1}$}
\def\sp#1{$^{#1}$}

\begin{document}

\draft

\wideabs{
\title{Neutron diffraction evidence of microscopic charge inhomogeneities in
the CuO$_2$ plane of 
superconducting La$\bf _{2-x}$Sr$\bf _x$CuO$\bf _4$ ($\bf 0\le x\le0.30$) }

\author{E. S. Bo\v zin,$^1$ G. H. Kwei,$^2$ H. Takagi$^3$ and S. J. L. 
Billinge$^1$}
\address{$^1$Department of Physics and Astronomy and Center for
Fundamental Materials Research, Michigan State University, 
East Lansing, MI 48824-1116.\\ 
$^2$Los Alamos National Laboratory, Los Alamos, NM 87545.\\
$^3$Department of Advanced Materials Science, University of Tokyo,
Hongo 7-3-1, Bunkyo-ku, Tokyo 113-8656, Japan}

\date{\today}

\maketitle

\begin{abstract}
We present local structural evidence supporting the presence of charge
inhomogeneities in the CuO$_2$ planes of underdoped
La$_{2-x}$Sr$_x$CuO$_4$.  High-resolution atomic pair distribution
functions have been obtained from neutron powder diffraction data over
the range of doping $0\le x\le 0.30$ at 10~K.  Despite the average
structure getting less orthorhombic we see a {\it broadening} of the
in-plane Cu-O bond distribution as a function of doping up to optimal
doping.  Thereafter the peak abruptly sharpens.  The peak broadening
can be well explained by a local microscopic coexistence of doped and
undoped material.  This suggests a crossover from a charge
inhomogeneous state at and below optimal doping to a homogeneous charge state
above optimal doping.  The strong response of the local structure to
the charge-state implies a strong electron-lattice coupling in these
materials.

\end{abstract}
}

There is mounting interest in the possibility that the charge distribution
in the CuO$_2$ planes of the high-temperature superconductors is 
microscopically inhomogeneous~\cite{egami;b;pphtsv96,tranq;n95} and 
that this has a bearing on the
high temperature superconductivity itself.  An inhomogeneous charge
distribution has been predicted 
theoretically%
~\cite{emery;pc93,zaane;prb89,phill;pmb99,bianc;ssc97,marki;jpcs97}.  
There is also mounting
experimental evidence that microscopic charge inhomogeneities exist
in particular cuprate samples.  The most compelling evidence for this
comes from the observation from neutron diffraction 
of stripes of localized charge in
La$_{2-x-y}$Nd$_{y}$Sr$_{x}$CuO$_{4}$~\cite{tranq;n95,tranq;prb96}.
However, such direct evidence for charge stripes has only been seen in
insulating compounds or in samples where this ordering competes with
superconductivity~\cite{tranq;prl97i}.  The evidence supporting the
presence of dynamic local charge stripe distributions in the bulk of
{\it superconducting} samples is based primarily on the observation of
incommensurate spin
fluctuations~\cite{mason;prl92,thurs;prb92,yamad;prb98,wakim;prb99}.
We present local structural evidence that supports the fact that the
charges are inhomogeneous in the underdoped and optimally doped region
of the La\sb{2-x}Sr\sb{x}CuO\sb{4} phase diagram, consistent with the
presence of charge domains or dynamic charge-stripes.
The charge distribution becomes homogeneous on crossing into the
overdoped region.  The clear observation of these effects in the local
structure also underscores the point that there is a strong
electron-lattice coupling, at least to particular distortion modes, in
these materials.  We note that the disappearance of the structural
distortions, and therefore the charge inhomogeneities, correlates with
the disappearance of the normal-state pseudogap~\cite{timus;rpp99} in
these materials.

The presence of charge inhomogeneities in the CuO$_2$ planes implies
profound consequences for the local structure.  It is well known that the
average Cu-O bond length changes as the charge state of copper changes.
Thus, the Cu-O bond in La$_{2-x}$Sr$_{x}$CuO$_{4}$ shortens from 1.904~\AA\
to 1.882~\AA\ as $x$ changes from 0 to 0.2~\cite{radae;prb94i} 
and the average copper charge changes
from $2+$ to $\sim 2.2+$.  This is a generic feature of variably dopable
HTS samples and comes about because the Cu-O bond is a covalent
anti-bonding state which is stabilized by removing electron density from it.
Clearly, if the doped charge in the CuO$_2$ planes is inhomogeneously
distributed, such
that some copper sites have more charge than others, a distribution of
in-plane Cu-O bond lengths will exist.  A high resolution measurement of the
in-plane Cu-O bond length {\it distribution} as a function of doping will 
therefore reveal the extent of charge inhomogeneities.

We have used the atomic pair distribution function (PDF) analysis of 
neutron powder diffraction data 
to measure accurate Cu-O bond length distributions with high resolution
for a series of La$_{2-x}$Sr$_{x}$CuO$_{4}$ samples with ($0\leq x\leq 0.3$). 
The ability of high-resolution PDF studies to reveal local bond-length
inhomogeneities which are not apparent in the average structure has been
clearly demonstrated~\cite{petko;prl99}.
The samples studied 
cover the range from undoped, through underdoped and optimally doped 
($x=0.15$) to
the overdoped regime.  We find that at 10~K
the mean-square width of the 
bond-length distribution, $\sigma^2$, increases approximately
linearly with $x$ until optimal doping above which it sharply decreases 
and returns
to the value of the undoped material by $x=0.25$.  This is strong evidence
for charge inhomogeneities in the under- and optimally-doped regimes 
as we discuss below.  This increase in bond-length distribution can be well
explained by a linear superposition of the local structures of undoped
and heavily doped material.

Samples of La$_{2-x}$Sr$_{x}$CuO$_{4}$ with $x=0.0$, 0.05, 0.1,
0.125, 0.15, 0.2, 0.25, and 0.3 were made using standard solid state
synthesis.  Mixtures of La\sb{2}O\sb{3}, SrCO\sb{3} and CuO were
reacted at temperatures between 900$^\circ$C and 1050$^\circ$C with
intermediate grindings, followed by an annealing step in air at
1100$^\circ$C for 100 hours and in oxygen at 800$^\circ$C for 100
hours.  The long anneals were carried out to ensure doping homogeneity
in the samples.  The lattice $c$-axis parameter is a sensitive
measure of oxygen stoichiometry~\cite{yamad;prb98}.  Its value was 
determined for each of our samples thru Rietveld refinement of the data.  It 
varies smoothly with $x$ as expected for stoichiometric samples
and lies on the $c$-axis vs $x$
curve of Radaelli {\it et al.}~\cite{radae;prb94i} within the scatter
of their data. 
Neutron powder diffraction data
were collected at 10~K on the Special Environment Powder
Diffractometer (SEPD) at the Intense Pulsed Neutron Source (IPNS) at
Argonne National Laboratory. Approximately 10g of finely powdered
sample was sealed in a cylindrical vanadium tube with He exchange gas.
The samples were cooled using a closed-cycle He refrigerator. The data
were corrected~\cite{billi;prb93} for experimental effects and
normalized to obtain the total structure function $S(Q)$ where $Q$
represents neutron momentum transfer.  The PDF, $G(r)$, is obtained by
a Fourier transform of the data according to $G(r) = {2\over
\pi}\int_0^{\infty} Q[S(Q)-1]\sin Qr\>dQ$.  PDFs from these samples
are shown in Figs. 5-8 of Ref.~\cite{bozin;prb99} and in
Fig.~\ref{fig;mix} of this paper.  The PDFs examined in this paper
used data over a range 0.7~\AA\sp{-1}$<Q<28$~\AA\sp{-1}.

\begin{figure}[!tb]
\begin{center}$\,$
\epsfxsize=2.8in
\epsfbox{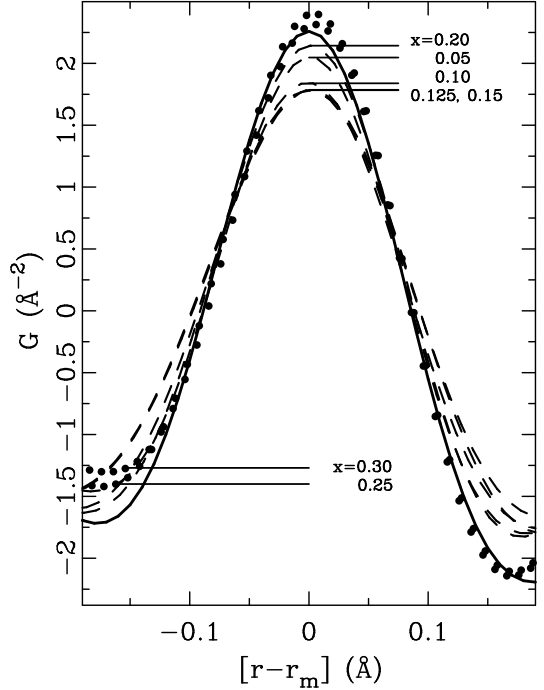}
\end{center}
\caption{PDF peak coming from the in-plane Cu-O bond for various doping
levels (solid line: undoped case; solid circles: overdoped cases; 
dashed lines: intermediate doping cases).  The peaks have been shifted 
so that their centers line up at $r_m = 1.91$~\AA .}
\protect\label{fig;firstpeak}
\end{figure}

\begin{figure}[!tb]
\begin{center}$\,$
\epsfxsize=2.8in
\epsfbox{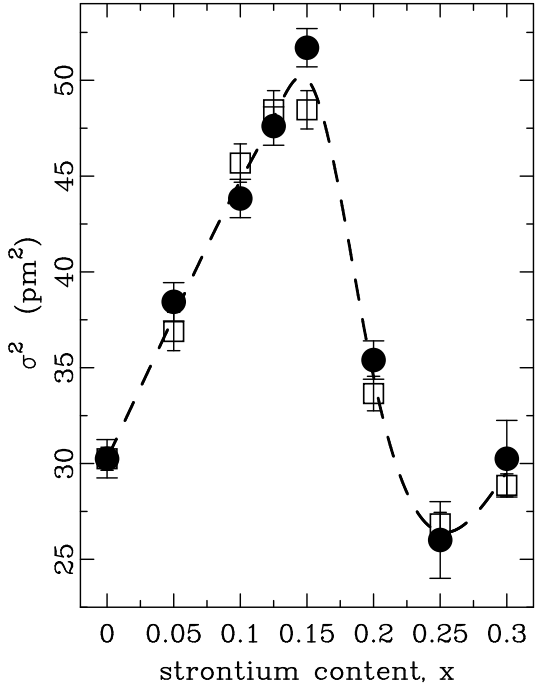}
\end{center}
\caption{Peak width of the in-plane Cu-O PDF peak as a function of 
doping obtained by fitting a Gaussian (solid circles). 
The data are plotted as $\sigma^2$ where $\sigma$ is the Gaussian 
standard deviation. The inverse peak-height-squared of 
the peaks in Fig.~1 scaled so the $x=0.0$ points line up is shown 
as open squares. The dashed line is a guide for the eye.}
\protect\label{fig;widthvx}
\end{figure}

\begin{figure}[!tb]
\begin{center}$\,$
\epsfxsize=2.8in
\epsfbox{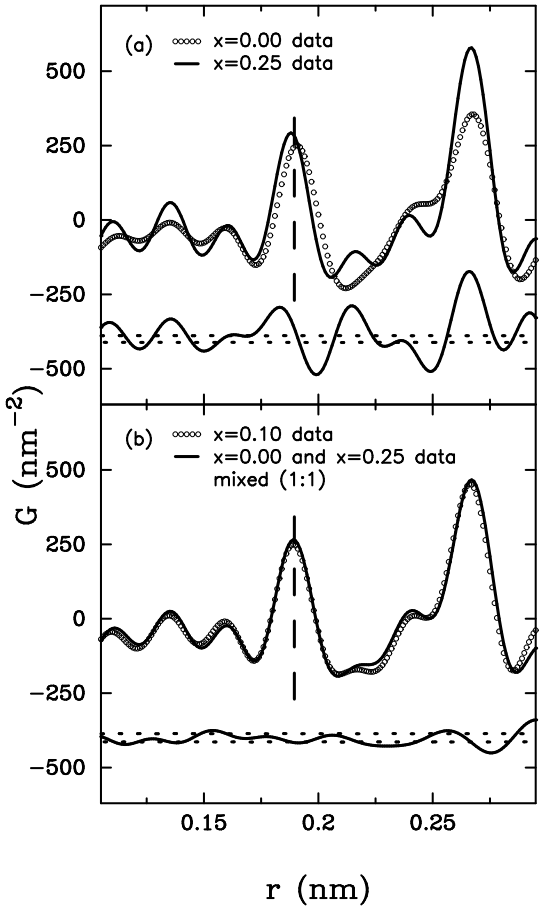}
\end{center}
\caption{(a) PDFs from the 10~K data from the $x=0.0$ (open circles) 
and $x=0.25$ (solid line) samples. The difference is plotted below. 
The dashed lines indicate expected uncertainties due to random errors. 
(b) PDF from $x=0.1$ data at 10~K (open circles). The solid line shows 
the PDF obtained by making a linear combination in a (1:1) ratio of 
the PDFs shown in (a).  See the text for details.}
\protect\label{fig;mix}
\end{figure}

We are interested in extracting the width of the distribution of
in-plane Cu-O bond-lengths.  This information is contained in the
width of the first PDF peak at $\sim 1.9$~\AA .  The peak-width comes
from the thermal and zero-point motion of the atoms plus any
bond-length distribution originating from other effects such as charge
inhomogeneities.  We can determine the latter by considering the PDF
peak width of this peak as a function of doping at 10~K.  Three
independent measures of the peak width all show that the width
increases significantly with doping up to $x=0.15$, beyond which the
peak quickly sharpens.  First, this behavior is evident by simply
looking at the data shown in Fig.~\ref{fig;firstpeak}.
This shows the low-$r$ region of the PDF around the 
$r=1.9$~\AA\ peak as a function of $x$. 
Since we want to compare the
relative peak widths (and heights), in this Figure the peaks have been 
shifted to
line up the peak centroids and rescaled slightly to ensure that the
integrated intensity of each peak is the same. 
It is clear that some of the peaks are significantly broader than others
with lower peak heights and less steeply sloping sides.  We have quantified
this by fitting the peak with a single Gaussian.  The Gaussian is first 
convoluted with the Fourier transform of the step function which was used
to terminate the 
data~\cite{billi;b;lsfd98}.
This accounts for any termination ripples in the
data introduced by the Fourier transform of the finite range data-set and
doesn't introduce any additional parameters into the fit. The 
fitting parameters were peak position,
scale-factor and width.  The baseline is set by the average number density
of the material and this was fixed at $\rho_0 = 0.07299$~\AA$^{-3}$.  
The results are 
shown in Fig.~\ref{fig;widthvx} as the solid circles.
The mean-square width of the distribution increases monotonically (and
almost linearly) with $x$ until $x=0.15$. Between 0.15 and 0.2 the
peak abruptly sharpens and returns to the width of the undoped sample
by $x=0.25$.  The same behavior can be obtained from the data in a
totally model-independent way.  If the integrated area of a Gaussian
is constant the height is inversely proportional to the width.  Thus,
the peak height, $h$, of the rescaled data shown in
Fig.~\ref{fig;firstpeak} should be inversely proportional to the
width.  The open squares in Fig.~\ref{fig;widthvx} show $C/h^2$ where
$h$ was determined directly from the peak maximum in the data and the
constant $C$ was chosen to make the $x=0.0$ points line up.  There is
excellent agreement lending confidence to the results from the
fitting.

We would like to discuss possible origins for these doping dependent
changes in Cu-O bond-length distribution.  First we rule out the
possibility that it simply comes from changes in the orthorhombicity
of the sample.  The Cu-O PDF peak first broadens smoothly with
increasing doping then dramatically sharpens at a composition close to
the LTO-HTT structural phase boundary.  This behavior does not reflect
the monotonic decrease in orthorhombicity of the average structure.
Indeed, in the overdoped region the PDF peak returns to the same
narrow width it had in the undoped material which has the largest
orthorhombic distortion of any of the samples.  We also note that the
in-plane Cu-O bonds are not expected to be sensitive to the
orthorhombic distortion.  They lie along the unit cell diagonals and
not along the unit cell edges in the orthorhombic unit cell.  In this
case an orthorhombic distortion will change the bond-length but will
not lead to two distinct bond-lengths. 
Next we show that the observed behavior cannot be explained by doping 
dependent changes in the octahedral tilts. 
The average~\cite{radae;prb94i} (and local)~\cite{bozin;prb99}
tilt amplitude monotonically decreases with increasing doping which
does not correlate with the behavior of the Cu-O bond length
distribution.  On the other hand, PDF peaks at higher-$r$ do sharpen
monotonically with increasing $x$ reflecting the reduced octahedral
tilt amplitude~\cite{bozin;prb99}.  We also rule out the idea that the
observed Cu-O peak broadening is an effect of size-effect disorder due
to doping since the peak sharpens dramatically 
above $x=0.2$ where the dopant induced disorder should be the
greatest.  We also note that size-effect dopant induced disorder is
expected to have a large effect on octahedral tilts and a small effect
on Cu-O bond lengths since the energy to change a Cu-O-Cu bond angle
is much less than the energy to stretch the short Cu-O covalent
bond. Furthermore, the extent of dopant induced tilt disorder is
relatively small in La\sb{2-x}Sr\sb{x}CuO\sb{4} as evidenced by the
observation that higher-$r$ PDF peaks sharpen on increased doping in
this material.  These peaks sharpen because of a decrease in both the
average orthorhombicity and octahedral tilt angle with increased
doping.  However, significant size-effect octahedral tilt disorder due
to the chemical dopants tends to counter this effect, as seen in
La\sb{2-x}Ba\sb{x}CuO\sb{4}~\cite{haske;unpub00,bozin;unpub00}.  We
can also rule out structural fluctuations associated with the HTT-LTO
transition.  First, these fluctuations will affect primarily
octahedral tilts and local orthorhombicity (the two order parameters
of this structural transition) and as we have discussed, the Cu-O bond
is expected to be quite insensitive to disorder in these parameters.
However, in addition, we would expect these fluctuations to be largest
when the structural phase transition temperature, T\sb{s}, is closest
to our measurement temperature.  These temperatures are closest for
the $x=0.2$ sample (T\sb{s}=60~K~\cite{Takag;prl92}, T\sb{meas}=10~K) 
and this sample exhibits a narrow distribution of Cu-O bond-lengths.  
The largest distribution of Cu-O bond-lengths is seen for $x=0.15$ where
T\sb{s}=180~K.

The observed
behavior of the Cu-O bond-length distribution is best explained by
the presence of charge inhomogeneities.  As we have described, the
charge-state of copper has a direct effect on the Cu-O bond length
with the bond-length decreasing with increasing doping.  Charge
inhomogeneities will, thus, give rise to a distribution of Cu-O bond
lengths.  Increased doping will result in more Cu-O bonds
being affected and therefore a larger measured effect in the PDF, as
observed.  Above optimal doping the PDF peak width abruptly sharpens
to its value in the undoped material. This is consistent with the idea
that in the overdoped region, the charge distribution in the Cu-O planes
is becoming homogeneous.
 
We now discuss independent evidence from the data which supports this
picture. 
In Fig.~\ref{fig;mix}(a) we show the low-$r$ region of the 
PDF from the $x=0.0$ and $x=0.25$
samples.  Referring to Fig.~\ref{fig;widthvx} we see that these two
data-sets have relatively {\it narrow} Cu-O bond-length distributions.
Furthermore, in Fig.~\ref{fig;mix}(a) it is apparent that the peak
position has shifted due to the change in the average Cu-O bond-length
with doping, as expected.  The difference curve below the data shows
that the two data-sets are quite different due to the significant
structural differences.  In Fig.~\ref{fig;mix}(b) we show the
intermediate $x=0.1$ data-set plotted as open circles.  This peak is
centered at a position shown by the dashed line which is intermediate
between the positions of the $x=0.0$ and $x=0.25$ data-sets.
Referring to Fig.~\ref{fig;widthvx} we see that the Cu-O bond-length
distribution is relatively {\it broad} at this composition.  Plotted on top
of the $x=0.1$ data-set in Fig.~\ref{fig;mix}(b) as the solid line is
the PDF obtained by taking a linear combination of the $x=0.0$ and
$x=0.25$ data-sets in the 1:1 ratio, without rescaling the data at
all.  The difference curve is shown below.  The good agreement clearly
demonstrates that the observed PDF peak position {\it and} broadening
of the $x=0.1$ data-set is entirely {\it consistent} with there being an
underlying bimodal bond-length distribution consistent with heavily
doped and undoped regions of the CuO\sb{2} plane.  We can infer from
this analysis that the difference in the bond-lengths is $\sim
0.024$~\AA\ which is the difference between the average bond-lengths of
the $x=0.0$ and $x=0.25$ samples.

Finally, we note that the PDFs from underdoped
La$_{2-x}$Sr$_{x}$CuO$_{4}$ are consistent with the presence of
CuO\sb{6} octahedral tilt disorder in the samples.  In an earlier
paper~\cite{bozin;prb99} we showed that the measured PDFs could be
well explained by a local structure which contains a mixture of large
and small octahedral tilts.  We note here that we also have evidence
from the PDFs for the presence of tilt-directional disorder in the
sense that there is a mixture of $\langle 100\rangle$ (``LTO'') and
$\langle 110\rangle$ (``LTT'') symmetry tilts present in the local
structure.  This will be reported in detail
elsewhere~\cite{bozin;unpub00}.

To summarize, we have presented evidence from neutron diffraction
data which strongly supports the idea that doped charge 
in the CuO$_2$ planes of superconducting 
La$_{2-x}$Sr$_{x}$CuO$_{4}$ for $0<x \leq 0.15$ and at 10~K
is inhomogeneous.  For doping levels of $x=0.2$ and above the
charge distribution in the Cu-O plane becomes homogeneous.  This presumably
reflects a crossover towards more fermi-liquid-like behavior in the
overdoped regime.

This work was supported financially by NSF through grant DMR-9700966.  SJB is 
a Sloan Research Fellow of the Alfred P. Sloan Foundation. 
GHK is funded at Los Alamos by the Department of Energy under Contract 
W-7405-ENG-36. 
The experimental data were collected at the IPNS at Argonne National
Laboratory. This facility is funded by the US 
Department of Energy under Contract W-31-109-ENG-38.

\end{document}